\documentclass[letterpaper,preprintnumbers,twocolumn,eqsecnum,superscriptaddress,aps,nofootinbib]{revtex4}
\usepackage{amssymb}\usepackage[centertags]{amsmath}\usepackage{txfonts}\usepackage{epsfig}\usepackage{bm}
\usepackage{color}\usepackage{graphicx,graphics}\usepackage{multirow}\usepackage{float}\usepackage{ulem}
\usepackage{hyperref}\usepackage{setspace}\usepackage{slashed}\usepackage{booktabs}
\hypersetup{colorlinks=true, citecolor=blue, linkcolor=red, filecolor=black,urlcolor=blue}
\allowdisplaybreaks[4]

\begin{document}

\title{Intrinsic asymmetries in semi-inclusive deeply inelastic scattering at the \\ Electron-Ion Collider}

\author{Weihua Yang}
\affiliation{College of Nuclear Equipment and Nuclear Engineering, Yantai University,\\ Yantai, Shandong 264005, China}

\author{Xinghua Yang~\footnote{yangxinghua@sdut.edu.cn (Corresponding author)}}
\affiliation{School of Physics and Optoelectronic Engineering, Shandong University of Technology,\\ Zibo, Shandong 255000, China}

\begin{abstract}
We calculate the neutral current jet production semi-inclusive deeply inelastic scattering process in this paper. Neutral current implies that interactions can be mediated by the photon, $Z^0$-boson and their interference. The initial electron is assumed to be polarized and then scattered off by a target particle with spin-1/2. Calculations are carried out up to twist-3 level in the quantum chromodynamics parton model by applying the collinear expansion formalism where multiple gluon scattering is taken into account and gauge links are obtained automatically. After obtaining the differential cross section, we introduce the definition of the intrinsic asymmetry. This quantity reveals the asymmetry in the distribution of the quark intrinsic transverse momentum. We find that these asymmetries can be expressed in terms of the transverse momentum dependent parton distribution functions and the electroweak couplings. As a result, our calculations provide a set of new quantities for analyzing the parton distribution functions and the electroweak couplings.  It is helpful to understand the hadronic weak interactions and strong interactions in the  deeply inelastic scattering process simultaneously.

\end{abstract}

\maketitle

\section{Introduction}

The use of leptons to probe the structure of the nucleon in deeply inelastic scattering (DIS) has achieved great success in the past decades. It will still play an important role in the future Electron-Ion Collider (EIC)~\cite{Accardi:2012qut,AbdulKhalek:2021gbh,Anderle:2021wcy} experiments.  One of the primary goals of the EIC is to explore the three-dimensional (3D) imaging of the nucleon or to measure the  transverse momentum dependent parton distribution functions (TMDs) over wide kinematic regions at high experimental precision. That is vital to understand the orbital motion, spin-orbital correlation as well as the spatial distribution of the parton in the quantum chromodynamics (QCD) bound state.  The EIC which has a large center-of-mass energy range also makes it possible for electroweak measurements through the neutral and charged current interactions, e.g., precision measurements of the weak mixing angle, parity violating asymmetries and charge asymmetries ~\cite{Boer:2011fh,Cahn:1977uu,Anselmino:1993tc,Prescott:1978tm,Prescott:1979dh,PVDIS:Jlab6,PVDIS:JLab12,Chen:2020ugq,Yang:2020qsk}.

The TMDs are usually extracted from the hadron production semi-inclusive DIS (SIDIS) data within the TMD formalism. Additionally, jet production SIDIS process attracted lots of attentions in recent years in extracting these TMDs  \cite{Song:2010pf,Song:2013sja,Wei:2016far,Gutierrez-Reyes:2018qez,Gutierrez-Reyes:2019vbx,Liu:2018trl,Liu:2020dct,Kang:2020fka,Arratia:2020ssx,H1:2021wkz}. Comparing to the hadron production SIDIS, the jet production one has two distinct features. First, jet production reaction does have simpler forms and not introduce extra uncertainties from fragmentation functions. This is helpful to improve the measurement accuracy. Second, the current region jet can be a direct probe of analyzing properties of the quark transverse momentum in the $\gamma^* N$ collinear frame. In this frame the transverse momentum of the virtual photon ($\vec q_\perp$) is zero. The transverse momentum of the jet ($\vec k'_\perp$) is equal to that of the incident quark ($\vec k_\perp$) if the higher order gluon radiations are neglected. %because of $\vec k'_\perp=\vec q_\perp+\vec k_\perp$.
Therefore,  the measurement of the jet can access to the information of the corresponding incident quark, even the correlation with the target nucleon spin.
Under this circumstance, we consider the neutral current jet production SIDIS process at the EIC energies to explore the transverse momentum properties of the quark in a nucleon. The neutral current here implies that interactions can be mediated by the photon, $Z^0$-boson and their interference. Semi-inclusive implies that a final current region jet is also measured in addition to the scattered lepton, i.e., the jet production SIDIS. The jet is simplified as a quark in our consideration. The initial electron is assumed to be polarized and then scattered off by a nucleon with spin-1/2.

Our calculations are carried out up to the leading order twist-3 (sub-leading power) level in the QCD parton model by applying the collinear expansion formalism \cite{Ellis:1982wd,Qiu:1990xxa,Liang:2006wp}. Higher twist effects are often significant for semi-inclusive reactions and/or TMD observables. Especially for the case of twist-3 corrections, they often lead to azimuthal asymmetries which are different from the leading twist ones~\cite{Mulders:1995dh,Bacchetta:2006tn}. Therefore, the studies of higher twist effects will give complementary or even direct access to the nucleon structures.
We calculate the twist-3 differential cross section of the jet production SIDIS process and introduce the definition of a new kind of asymmetry. This quantity named as intrinsic asymmetry reveals the asymmetry in the distribution of the quark transverse momentum in a nucleon. We obtain eight $S_T$-independent asymmetries and four $S_T$-dependent asymmetries with well definitions.  We find that these asymmetries can be expressed in terms of the TMDs and the electroweak couplings. As a result, our calculations provide a set of new quantities for analyzing these corresponding TMDs and the electroweak couplings.  It is helpful to understand the hadronic weak interactions and strong interactions in the deeply inelastic scattering process simultaneously.

The rest of this paper is organized as follows. In Sec.~\ref{sec:formalism}, we present the formalism of jet production SIDIS process and calculate the hadronic tensor at the leading order twist-3 level in terms of the TMDs in the parton model.  In Sec.~\ref{sec:crosssection}, we calculate the differential cross section and introduce the definition of the intrinsic asymmetry. Detailed expressions and numerical results are also shown there.
Finally, a brief summary is given in Sec.~\ref{sec:summary}.

\section{The process and the hadronic tensor} \label{sec:formalism}

\subsection{The formalism}

We consider the current region jet production SIDIS process at EIC energies. To be explicit, this process can be labeled as
\begin{align}
e^-(l,\lambda_e) + N(p,S) \rightarrow e^-(l^\prime) + q(k^\prime) + X,
\end{align}
where $\lambda_e$ is the helicity of the initial electron with momentum $l$. $N$ can be a nucleon with momentum $p$, spin-1/2. $q$ denotes a quark which corresponds to a jet of hadrons observed in experiments. In this paper, we consider the case of the electron scattered off a spin-1/2 target with the neutral current interaction at the tree level of electroweak theory, i.e., the exchange of a virtual photon $\gamma^*$ or a $Z^0$ boson.
The standard variables used in this paper for the SIDIS are
\begin{align}
  x=\frac{Q^2}{2 p\cdot q}, \  y=\frac{p\cdot q}{p \cdot l},\  s=(p+l)^2,
\label{eq:SIDIS-var}
\end{align}
where $Q^2 = -q^2=-(l-l')^2$.
%\begin{align}
%d\sigma = \sum_X \frac{|{\cal M}|^2}{2s} (2\pi)^4 \delta^4 (l + p - l^\prime - k^\prime -p_X) \frac{d^3 l^\prime} {(2\pi)^3 2E_{l^\prime}} \frac{d^3 k^\prime}{(2\pi)^3 2E_{k^\prime}}.
%\end{align}
%where $\sum_X = \Pi_X \int \frac{d^3 p_X}{(2\pi)^3 2E_X}$. At high energy collider like EIC, the exchange of a $Z^0$ boson between the electron and the ion can be relevant. The scattering amplitude is then given by
%\begin{align}
%\mathcal{M} = {\cal M}_{\gamma} + {\cal M}_Z,
%\end{align}
%in which
%\begin{align}
%& {\cal M}_{\gamma} = \bar u_{s^\prime}(l^\prime) (-ie\gamma_\mu) u_{s}(l)  \frac{-ig^{\mu\nu}}{q^2} \langle k^\prime;X| J_\nu^{em}(0) |p,S \rangle, \nonumber\\
%& {\cal M}_Z = \bar u_{s^\prime}(l^\prime) \left(\frac{-ie\Gamma_\mu^e}{\sin2\theta_W}\right) u_{s}(l) \frac{-i\left(g^{\mu\nu} - q^\mu q^\nu/M_Z^2\right)}{q^2 - M_Z^2 + i\Gamma_Z M_Z} \langle k^\prime;X| J_\nu^{w}(0) |p,S \rangle,
%\end{align}
%with $J_\nu^{em}(0)$ and $J_\nu^{w}(0)$ the electromagnetic and weak current. In the QCD parton model
%\begin{align}
%& {\cal M}_{\gamma} = \bar u_{s^\prime}(l^\prime) (-ie\gamma_\mu) u_{s}(l)  \frac{-ig^{\mu\nu}}{q^2} \langle k^\prime;X| (-ie_q e) \bar\psi(0) \gamma_\nu \psi(0) |p,S \rangle, \nonumber\\
%& {\cal M}_Z = \bar u_{s^\prime}(l^\prime) \left(\frac{-ie\Gamma_\mu^e}{\sin2\theta_W}\right) u_{s}(l) \frac{-i\left(g^{\mu\nu} - q^\mu q^\nu/M_Z^2\right)}{q^2 - M_Z^2 + i\Gamma_Z M_Z} \langle k^\prime;X| \left(\frac{-ie}{\sin2\theta_W}\right) \bar\psi(0) \Gamma_\nu^q \psi(0) |p,S \rangle,
%\end{align}
The differential cross-section is written as
\begin{align}
  d\sigma = \frac{\alpha_{\rm em}^2}{sQ^4}A_r L^r_{\mu\nu}(l,\lambda_e, l^\prime)W_r^{\mu\nu}(q,p,S,k^\prime)\frac{d^3 l^\prime d^3 k^\prime}{(2\pi)^32E_{l^\prime} E_{k^\prime}}. \label{f:crosssec}
\end{align}
$\alpha_{\rm em}$ is the fine structure constant. The symbol $r$ can be $\gamma\gamma$, $ZZ$ and $\gamma Z$, for electromagnetic (EM), weak and interference terms, respectively.
A summation over $r$ in Eq. (\ref{f:crosssec}) is understood, i.e., the total cross section is given by
\begin{align}
  %\frac{d\sigma}{dx dy d\psi d^2 k_\perp^\prime} = \frac{d\sigma^{ZZ} + d\sigma^{\gamma Z} + d\sigma^{\gamma\gamma}}{dx dy d\psi d^2 k_\perp^\prime}.
  d\sigma = d\sigma^{ZZ} + d\sigma^{\gamma Z} + d\sigma^{\gamma\gamma}.
\end{align}
$A_r$'s are defined as
\begin{align}
& A_{\gamma\gamma} = e_q^2, \nonumber\\
& A_{ZZ} = \frac{Q^4}{\left[(Q^2+M_Z^2)^2 + \Gamma_Z^2 M_Z^2 \right] \sin^4 2\theta_W} \equiv \chi, \nonumber\\
& A_{\gamma Z} = \frac{2e_q Q^2 (Q^2+M_Z^2)}{\left[(Q^2+M_Z^2)^2 + \Gamma_Z^2 M_Z^2 \right] \sin^2 2\theta_W}\equiv \chi_{int}.
\end{align}
The leptonic tensors for the EM, weak and interference interactions are respectively given by
\begin{align}
%& L_{\gamma\gamma}^{\mu\nu}(l,\lambda_e, l^\prime) ={\rm Tr} \sum_{s^\prime}\left[ u_{s^\prime}(l^\prime)\bar u_{s^\prime}(l^\prime) \gamma^\nu u_{s}(l)\bar u_{s}(l) \gamma^\mu \right]
 &L^{\gamma\gamma}_{\mu\nu}(l,\lambda_e, l^\prime)= 2\left[ l_\mu l^\prime_\nu + l_\nu l^\prime_\mu - (l\cdot l^\prime)g_{\mu\nu}  \right] + 2i\lambda_e \varepsilon_{\mu\nu l l^\prime}, \\
 & L^{ZZ}_{\mu\nu}(l,\lambda_e, l^\prime) =(c_1^e - c_3^e \lambda_e)L^{\gamma\gamma}_{\mu\nu}(l,\lambda_e, l^\prime), \\
 & L^{\gamma Z}_{\mu\nu}(l,\lambda_e, l^\prime)=(c_V^e - c_A^e \lambda_e) L^{\gamma\gamma}_{\mu\nu}(l,\lambda_e, l^\prime),
\end{align}
where $c_1^e = (c_V^e)^2 + (c_A^e)^2$ and $c_3^e = 2 c_V^e c_A^e$.
$c_V^e$ and $c_A^e$ are defined in the weak interaction current
$J_\mu (x)=\bar \psi(x)\Gamma_\mu\psi(x)$ with $\Gamma^e_\mu= \gamma_\mu (c_V^e - c_A^e \gamma^5)$.
Similar notations are also used for quarks where the superscript $e$ is replaced by $q$.
The hadronic tensors are given by
\begin{align}
  W_{\gamma\gamma}^{\mu\nu}&(q,p,S,k^\prime) = \sum_X (2\pi)^3 \delta^4(p + q - k^\prime - p_X)\nonumber\\
 &\quad \times\langle p,S |J_{\gamma\gamma}^\mu(0)|k^\prime;X\rangle \langle k^\prime;X |J_{\gamma\gamma}^\nu(0)| p,S \rangle , \\
  W_{ZZ}^{\mu\nu}&(q,p,S,k^\prime) = \sum_X (2\pi)^3 \delta^4(p + q - k^\prime - p_X)\nonumber\\
 &\quad \times\langle p,S | J_{ZZ}^\mu(0)|k^\prime;X\rangle \langle k^\prime;X | J_{ZZ}^\nu(0) | p,S \rangle , \\
  W_{\gamma Z}^{\mu\nu}&(q,p,S,k^\prime) = \sum_X (2\pi)^3 \delta^4(p + q - k^\prime - p_X)\nonumber\\
 &\quad \times\langle p,S |J_{ZZ}^\mu(0)|k^\prime;X\rangle \langle k^\prime;X | J_{\gamma\gamma}^\nu(0)| p,S \rangle ,
\end{align}
where $J_{\gamma\gamma}^\mu(0) =\bar\psi(0) \gamma^\mu \psi(0)$, $ J_{ZZ}^\mu(0) =\bar\psi(0) \Gamma^\mu_q \psi(0)$ with $\Gamma^\mu_q = \gamma^\mu(c_V^q - c_A^q \gamma_5)$.
It is convenient to consider the  $k_\perp^\prime$-dependent cross section, i.e.,
\begin{align}
  d\sigma = \frac{\alpha_{\rm{em}}^2}{sQ^4}A_{r} L^r_{\mu\nu}(l,\lambda_e, l^\prime)W_r^{\mu\nu}(q,p,S,k_\perp^\prime) \frac{d^3 l^\prime d^2 k_\perp^\prime}{E_{l^\prime}}, \label{f:crosssection}
\end{align}
where $k^{\prime}_z$ has been integrated and the integrated hadronic tensor is given by
\begin{align}
W_r^{\mu\nu}(q,p,S,k_\perp^\prime) = \int \frac{dk_z^\prime}{(2\pi)^3 2E_{k^\prime}} W_{r}^{\mu\nu}(q,p,S,k^\prime).  \label{f:haint}
\end{align}

In terms of the variables in Eq. (\ref{eq:SIDIS-var}), we can rewrite the cross section as
\begin{align}
\frac{d\sigma}{dx dy d\psi d^2 k_\perp^\prime} = \frac{y \alpha_{\rm em}^2}{2 Q^4}  A_{r}
L_{\mu\nu}^r(l,\lambda_e, l^\prime)W_r^{\mu\nu}(q,p,S,k_\perp^\prime), \label{f:intcross}
\end{align}
by using $d^3 l^\prime /2E_{l^\prime} \approx  y s dx d y d\psi/4$. Here $\psi$ is the azimuthal angle of $\vec l^\prime$ around $\vec l$.
%\begin{align}
%\frac{d^3 l^\prime}{2E_{l^\prime}} = \frac{y (s-M^2)}{4} dx d y d\psi \approx \frac{y s}{4} dx d y d\psi,
%\end{align}

\subsection{The hadronic tensor }

\begin{figure}
\centering
\includegraphics[width= 0.9\linewidth]{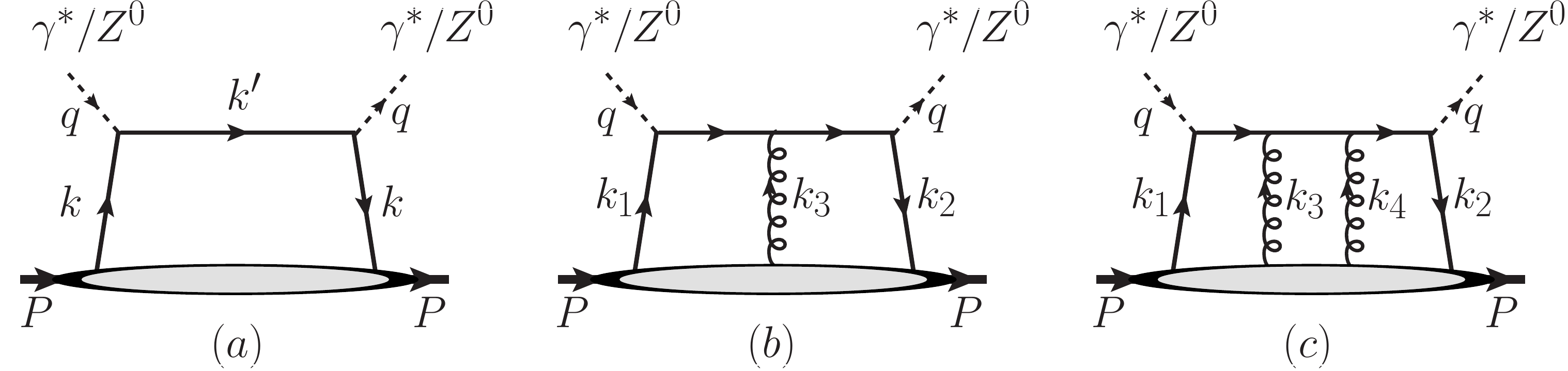}
\caption{The first few diagrams of the Feynman diagram series with exchange of $j$ gluons, where $j=0,~1$ and $2$ for diagrams $(a)$, $(b)$ and $(c)$ respectively.}
\label{fig:CoExp}
\end{figure}

At the tree level without higher order gluon radiations, the leading twist hadronic tensor gets contributions from the handbag diagram, see the first diagram (a) in Fig. \ref{fig:CoExp}. For higher twist contributions, multiple gluon scattering diagrams should be included, e.g.,
we consider the first two diagrams in Fig. \ref{fig:CoExp} at twist-3 level. Correspondingly, both the quark-quark and quark-gluon-quark correlation functions (denoted by the shaded regions) contribute to the hadronic tensor, they are defined as
\begin{align}
  \hat \phi^{(0)}(k,p,S) =& \int \frac{d^4y}{(2\pi)^3}e^{ikz}\langle p,S|\bar \psi(0)\psi(y) |p,S \rangle, \label{f:phi0}\\
  \hat \phi^{(1)}_\rho(k_1,k_2,p,S) =& \int \frac{d^4y}{(2\pi)^3}\frac{d^4z}{(2\pi)^3}e^{ik_1z+ik_2(y-z)}\nonumber\\
  &\times \langle p,S |\bar \psi(0)g A_\rho(z)\psi(y) |p,S \rangle, \label{f:phi1}
\end{align}
where $A_\rho$ is the gluon field. We can see that correlation functions in Eqs. (\ref{f:phi0})-(\ref{f:phi1}) are not gauge invariant. To obtain the gauge invariant forms, we use the collinear expansion method which was introduced decades ago for DIS \cite{Ellis:1982wd,Qiu:1990xxa} and then extended to the SIDIS \cite{Liang:2006wp}.
At the tree level, in order to calculate the hadronic tensor in the collinear expansion formalism, we need to consider the contributions from the series of diagrams shown in Fig.~\ref{fig:CoExp}, i.e., the multiple gluon scattering contributions. Detailed derivations can be found in refs. \cite{Song:2010pf,Song:2013sja,Liang:2006wp}, we do not repeat them in this paper for simplicity.

After collinear expansion, the hadronic tensor is expressed in terms of
the gauge-invariant quark-quark and quark-gluon-quark correlation functions and corresponding calculable hard parts at the twist-3 level, i.e.,
\begin{align}
W_{r,\mu\nu} (q,p,S,k^\prime) = \sum_{j,c} \tilde W_{r,\mu\nu}^{(j,c)} (q,p,S,k^\prime),
\end{align}
where $j$ denotes the number of exchanged gluons and $c$ denotes different cuts. After integration over $k_z^\prime$,  $\tilde W_{r,\mu\nu}^{(j,c)}$'s are simplified as
\begin{align}
& \tilde{W}_{r,\mu\nu}^{(0)}(q,p,S,k_\perp^\prime) = \frac{1}{2}{\rm Tr}\left[\hat{h}_{r,\mu \nu}^{(0)} \hat{\Phi}^{(0)}(x,k_\perp)\right] , \label{f:W0munu}\\
& \tilde{W}_{r,\mu\nu}^{(1, L)}(q,p,S,k_\perp^\prime) = \frac{1}{4 p \cdot q}{\rm Tr}\left[\hat{h}_{r,\mu\nu}^{(1) \rho} \hat{\Phi}_{\rho}^{(1)}(x,k_\perp)\right] \label{f:W1Lmunu}
\end{align}
up to the relevant twist-3 level. They correspond to Eq. (\ref{f:haint}). The hard parts $h_r$'s are
\begin{align}
& \hat h_{\gamma\gamma,\mu\nu}^{(0)} = \gamma_{\mu} \slashed n \gamma_{\nu} / p^+, \qquad \hat{h}_{\gamma\gamma,\mu\nu}^{(1) \rho}=\gamma_{\mu} \slashed{\bar n} \gamma_{\perp}^{\rho} \slashed n \gamma_{\nu}, \\
%%%%%%
& \hat h_{ZZ,\mu\nu}^{(0)} = \Gamma_{\mu}^q \slashed n \Gamma_{\nu}^q / p^+, \qquad \hat{h}_{ZZ,\mu\nu}^{(1) \rho}=\Gamma_{\mu}^q \slashed{\bar n} \gamma_{\perp}^{\rho} \slashed n \Gamma_{\nu}^q, \\
%%%%%%
& \hat h_{\gamma Z,\mu\nu}^{(0)} = \Gamma_{\mu}^q \slashed n \gamma_{\nu} / p^+, \qquad \hat{h}_{\gamma Z,\mu\nu}^{(1) \rho}=\Gamma_{\mu}^q \slashed{\bar n} \gamma_{\perp}^{\rho} \slashed n \gamma_{\nu}.
\end{align}

The gauge-invariant operator definitions  of the quark-quark and quark-gluon-quark correlation functions are defined as
\begin{align}
  \hat{\Phi}^{(0)}\left(x, k_{\perp}\right) =& \int \frac{p^{+} d y^{-} d^{2} y_{\perp}}{(2 \pi)^{3}} e^{i x p^{+} y^{-}-i \vec{k}_{\perp} \cdot \vec{y}_{\perp}} \nonumber\\
  &\times \langle p,S|\bar{\psi}(0) {\cal L}(0, y) \psi(y)| p,S\rangle, \\
  \hat{\Phi}_{\rho}^{(1)}\left(x, k_{\perp}\right) =& \int \frac{p^{+} d y^{-} d^{2} y_{\perp}}{(2 \pi)^{3}} e^{i x p^+ y^- - i \vec{k}_\perp \cdot \vec{y}_\perp}\nonumber\\
  &\times  \langle p,S| \bar{\psi}(0) D_{\perp \rho}(0) {\cal L}(0, y) \psi(y)| p,S\rangle,
\end{align}
where $D_\rho(y) = -i\partial_\rho + g A_\rho(y)$ is the covariant derivative.
${\cal L}(0, y)$ is the gauge link obtained from the collinear expansion procedure, which guarantees the gauge invariance of these correlation functions.

The quark-quark and quark-gluon-quark correlation functions are $4\times 4$ matrices in Dirac space which can be decomposed in terms of the Dirac Gamma-matrices and coefficient functions.
In the jet production SIDIS process $e^-N\to e^- q X$, where the fragmentation is not considered, only the chiral even PDFs are involved. Because there is no spin flip.
Therefore, we only need to consider the $\gamma^\alpha$- and the $\gamma^\alpha\gamma^5$-terms in the decomposition of these correlation functions.
We have
\begin{align}
& \hat \Phi^{(0)}= \frac{1}{2}\left[\gamma^\alpha \Phi^{(0)}_\alpha + \gamma^\alpha\gamma_5 \tilde\Phi^{(0)}_\alpha \right], \\
& \hat \Phi_\rho^{(1)}= \frac{1}{2}\left[\gamma^\alpha \Phi_{\rho\alpha}^{(1)} + \gamma^\alpha\gamma_5 \tilde\Phi_{\rho\alpha}^{(1)} \right] .
\end{align}
The TMDs are defined through the decomposition of the correlation functions or the coefficient functions. Following the convention in ref.~\cite{Wei:2016far}, we have
\begin{align}
  \Phi^{(0)}_\alpha &=p^+ \bar n_\alpha\Bigl(f_1-\frac{k_\perp \cdot \tilde S_T}{M}f^\perp_{1T}  \Bigr) +k_{\perp\alpha} f^\perp  \nonumber\\
  &- M\tilde S_{T\alpha}f_T - \lambda_h \tilde k_{\perp\alpha} f_L^\perp  -\frac{k_{\perp\langle\alpha}k_{\perp\beta\rangle}}{M}  \tilde S_T^\beta f_T^\perp  ,
\label{eq:Xi0Peven}\\
%%%%%%%%%%%%%%%%%%%%%%%%%%%%
  \tilde\Phi^{(0)}_\alpha &=p^+\bar n_\alpha\Bigl(-\lambda_hg_{1L}+\frac{k_\perp\cdot S_T}{M}g^\perp_{1T}\Bigr)-\tilde k_{\perp\alpha}  g^\perp\nonumber\\
  &- M S_{T\alpha}g_T -\lambda_h k_{\perp\alpha} g_L^\perp + \frac{k_{\perp\langle\alpha}k_{\perp\beta\rangle}}{M}  S_T^\beta g_T^\perp .
\label{eq:Xi0Podd}
\end{align}
Here $\tilde A^\alpha_\perp =\varepsilon_\perp^{\alpha A}=\varepsilon_\perp^{\alpha \beta}A_{\perp\beta}$, $A$ can be $k_\perp$ or $S_T$. $k_{\perp\langle\alpha}k_{\perp\beta\rangle}= k_{\perp\alpha}k_{\perp\beta}-g_{\perp\alpha\beta}k_\perp^2/2$.
For the quark-gluon-quark correlation function, we have
\begin{align}
  \Phi^{(1)}_{\rho\alpha}&=p^+\bar n_\alpha\Bigl[k_{\perp\rho} f^\perp_d- M\tilde S_{T\rho}f_{dT}  -\lambda_h \tilde k_{\perp\rho} f_{dL}^\perp \nonumber\\
  & -\frac{k_{\perp\langle\rho}k_{\perp\beta\rangle}}{M} \tilde S_T^\beta f_{dT}^\perp \Bigr], \label{eq:Xi1Peven} \\
%%%%%%%%%%%%%%%%%%%%%%%%%%%%%%%
  \tilde \Phi^{(1)}_{\rho\alpha}&=ip^+\bar n_\alpha\Bigl[\tilde k_{\perp\rho}g^\perp_d+ MS_{T\rho}g_{dT}+\lambda_h k_{\perp\rho} g_{dL}^\perp  \nonumber\\
  & - \frac{k_{\perp\langle\rho}k_{\perp\beta\rangle}}{M} S_T^\beta g_{dT}^\perp  \Bigr],
\label{eq:Xi1Podd}
\end{align}
where a subscript $d$ is used to denote TMDs defined via the quark-gluon-quark correlation function or coefficient functions.

In fact, not all twist-3 TMDs  shown in Eqs.~(\ref{eq:Xi0Peven})-(\ref{eq:Xi1Podd}) are independent. We can use the QCD equation of motion $\slashed D\psi=0$ to obtain the following equations to eliminate TMDs which are not independent, i.e.,
\begin{align}
 x p^{+} \Phi^{(0) \rho} &=-g_{\perp}^{\rho \sigma} \operatorname{Re} \Phi_{\sigma+}^{(1)}-\varepsilon_{\perp}^{\rho \sigma} \operatorname{Im} \tilde{\Phi}_{\sigma+}^{(1)}, \label{eq:eom1}\\
 x p^{+} \tilde{\Phi}^{(0) \rho} &=-g_{\perp}^{\rho \sigma} \operatorname{Re} \tilde{\Phi}_{\sigma+}^{(1)}-\varepsilon_{\perp}^{\rho \sigma} \operatorname{Im} \Phi_{\sigma+}^{(1)}.\label{eq:eom2}
\end{align}
By inserting Eqs.~(\ref{eq:Xi0Peven})-(\ref{eq:Xi1Podd}) into Eqs.~(\ref{eq:eom1}) and (\ref{eq:eom2}), we obtain the relationships between the twist-3 TMDs defined via the quark-quark correlation function and those defined via the quark-gluon-quark correlation function.
They can be written in a unified form, i.e.,
\begin{align}
f_{d S}^{K}-g_{d S}^{K}=-x\left(f_{S}^{K}-i g_{S}^{K}\right),\label{f:formulaEOM}
\end{align}
where $K$ can be $\perp$ and $S$ can be $L$ and $T$ whenever applicable.

Substituting the hard parts and the corresponding TMDs into Eqs. (\ref{f:W0munu}) and (\ref{f:W1Lmunu}) and using Eq. (\ref{f:formulaEOM}) to eliminating the independent TMDs  gives the complete twist-3 hadronic tensor:
\begin{widetext}
\begin{align}
\tilde W^{\mu\nu} =&
-\left( c_1^q g_{\perp}^{\mu\nu} + ic_3^q \varepsilon_{\perp}^{\mu\nu} \right) \Bigl(f_1-\frac{k_\perp \cdot \tilde S_T}{M}f^\perp_{1T} \Bigr)- \left( c_3^q g_{\perp}^{\mu\nu} + ic_1^q \varepsilon_{\perp}^{\mu\nu} \right) \Bigl(-\lambda_hg_{1L}+\frac{k_\perp\cdot S_T}{M}g^\perp_{1T} \Bigr) \nonumber \\
+&\frac{1}{(p \cdot q)} \Bigg\{\Bigl[ c_1^q k_\perp^{\{\mu} \bar q^{\nu\}} + ic_3^q \tilde k_\perp^{[\mu} \bar q^{\nu]} \Bigr] f^\perp- \Bigl[ c_1^q \tilde k_\perp^{\{\mu} \bar q^{\nu\}} - ic_3^q k_\perp^{[\mu} \bar q^{\nu]} \Bigr] \lambda_h f_L^\perp - \Bigl[ c_1^q \tilde S_T^{\{\mu} \bar q^{\nu\}} - ic_3^q S_T^{[\mu} \bar q^{\nu]} \Bigr] M f_T \nonumber\\
%%%%%%
&- \Biggl[ c_1^q \left( \frac{k_\perp\cdot \tilde S_T}{M} k_\perp^{\{\mu} \bar q^{\nu\}} - \frac{k_\perp^2}{2M} \tilde S_T^{\{\mu} \bar q^{\nu\}} \right) + ic_3^q \left(\frac{k_\perp\cdot S_T}{M} k_\perp^{[\mu} \bar q^{\nu]} - \frac{k_\perp^2}{2M}S_T^{[\mu} \bar q^{\nu]} \right) \Biggr] f_T^\perp \nonumber\\
%%%%%%
%%%%%%%%%%%%%%%%%%%%%%%%%%%
%%%%%%
&- \Bigl[ c_3^q \tilde k_\perp^{\{\mu} \bar q^{\nu\}} - ic_1^q k_\perp^{[\mu} \bar q^{\nu]}\Bigr]  g^\perp - \Bigl[ c_3^q k_\perp^{\{\mu} \bar q^{\nu\}} + ic_1^q \tilde k_\perp^{[\mu} \bar q^{\nu]}\Bigr] \lambda_h g_L^\perp- \Bigl[ c_3^q S_T^{\{\mu} \bar q^{\nu\}} + ic_1^q \tilde S_T^{[\mu} \bar q^{\nu]} \Bigr] M g_T \nonumber\\
%%%%%%
&+ \Biggl[ c_3^q \left( \frac{k_\perp\cdot S_T}{M} k_\perp^{\{\mu} \bar q^{\nu\}} - \frac{k_\perp^2}{2M} S_T^{\{\mu} \bar q^{\nu\}} \right)  + ic_1^q \left(\frac{k_\perp\cdot S_T}{M} \tilde k_\perp^{[\mu} \bar q^{\nu]} - \frac{k_\perp^2}{2M} \tilde S_T^{[\mu} \bar q^{\nu]} \right) \Biggr] g_T^\perp \Bigg\},  \label{f:Wt3munu}
\end{align}
\end{widetext}
where $\bar q^\mu = q^\mu + 2xp^\mu$.
The first line in Eq. (\ref{f:Wt3munu}) is the leading twist part while the other lines give the twist-3 part.
From $q\cdot\bar q = q\cdot k_\perp = 0$ and $q\cdot S_T =0$, we see clearly that the full twist-3 hadronic tensor satisfies current conservation, $q_\mu \tilde W^{\mu\nu}_{t3} = q_\nu \tilde W^{\mu\nu}_{t3} = 0$.

\section{The results up to twist-3} \label{sec:crosssection}

\subsection{The differential cross section}

\begin{figure}[ht]
\centering
\includegraphics[width= 0.9\linewidth]{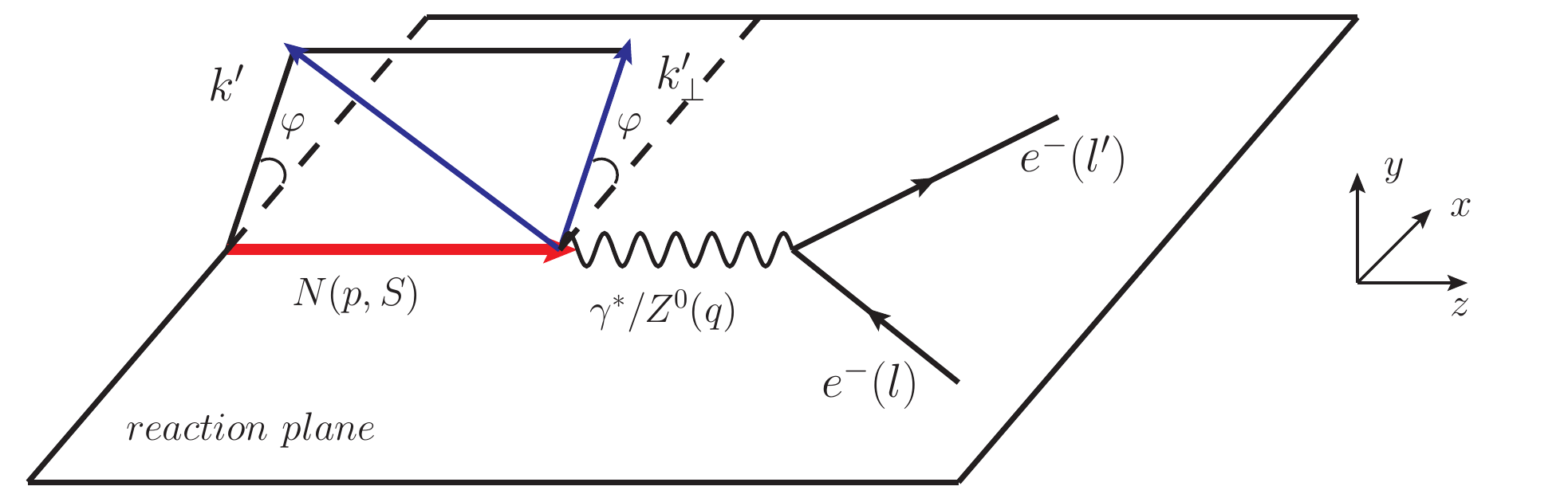}
\caption{Illustrating of the jet production SIDIS in the $\gamma^*N$ collinear frame. Momenta are labeled in the parentheses. Leptons are in the $x-z$ plane or the reaction plane. }
\label{fig:ins}
\end{figure}

\begin{figure}
\centering
\includegraphics[width= 0.6\linewidth]{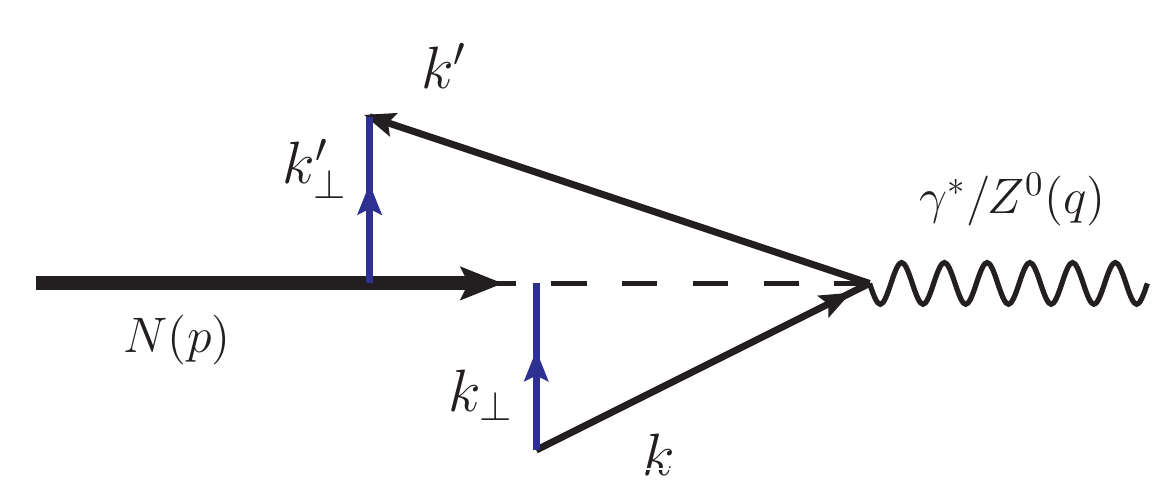}
\caption{Illustrating of the quark (jet) transverse momentum in the $\gamma^*N$ collinear frame.}
\label{fig:kperp}
\end{figure}

In order to calculate the differential cross section, we choose the $\gamma^*N$ collinear frame, see Fig. \ref{fig:ins}, in which the momenta related to this SIDIS process take the following forms:
\begin{align}
& p^\mu = \left(p^+,0,\vec 0_\perp \right), \nonumber\\
& l^\mu = \left( \frac{1-y}{y}xp^+, \frac{Q^2}{2xyp^+}, \frac{Q\sqrt{1-y}}{y},0 \right),\nonumber\\
& q^\mu = \left( -xp^+, \frac{Q^2}{2xp^+}, \vec 0_\perp \right), \nonumber\\
& k_\perp^{\prime\mu} = k_\perp^\mu = |\vec k_\perp| \left( 0,0, \cos\varphi, \sin\varphi \right).
\end{align}
Here $ k_\perp^{\prime\mu}$ denotes the transverse momentum vector of the jet and $ k_\perp^{\mu}$ denotes that of the quark in a nucleon. They are equal to each other in this frame, see Fig. \ref{fig:kperp}.  We do not distinguish them in the following discussions.
And the transverse vector polarization is parameterized as
\begin{align}
& S_T^\mu = |S_T| \left( 0,0, \cos\varphi_S, \sin\varphi_S \right).
\end{align}
We define the following functions of $y$ which will be often used:
\begin{align}
& A(y) = y^2-2y+2, \nonumber\\
& B(y) = 2(2-y)\sqrt{1-y}, \nonumber\\
& C(y) = y(2-y), \nonumber\\
& D(y) = 2y\sqrt{1-y}.
%& E(y) = 2(1-y).
\end{align}

It is convenient to divide the differential cross section into the leading twist and twist-3 parts.
Substituting the leading twist part of the hadronic tensor in Eq.~(\ref{f:Wt3munu}) and the leptonic tensor into Eq.~(\ref{f:intcross}) yields the leading twist cross section.
Here, we give the expressions explicitly for the weak interaction part
\begin{align}
  \frac{d\sigma_{t2}^{ZZ}}{dx dy d\psi d^2 k_\perp^\prime}& =\frac{\alpha_{\rm em}^2 \chi}{y Q^2}\bigg\{ \left[T_0^q(y) -\lambda_e \tilde T_0^q(y) \right] f_1 \nonumber\\
  &- \left[\tilde T_1^q(y) -\lambda_e T_1^q(y) \right] \lambda_h g_{1L} \nonumber\\
  &+|S_T|k_{\perp M}\Big[\sin(\varphi-\varphi_S) \bigl(T_0^q(y) -\lambda_e \tilde T_0^q(y) \bigr) f^\perp_{1T}\nonumber\\
  &-\cos(\varphi-\varphi_S) \bigl(\tilde T_1^q(y) -\lambda_e T_1^q(y) \bigr)g^\perp_{1T}\Big]
\bigg\},
\end{align}
where we have defined $k_{\perp M} = |\vec k_\perp|/M$, and
\begin{align}
  & T_0^q(y) = c_1^e c_1^q A(y) + c_3^e c_3^q C(y), \nonumber\\
  & \tilde T_0^q(y) = c_3^e c_1^q A(y) + c_1^e c_3^q C(y), \nonumber\\
  & T_1^q(y) = c_3^e c_3^q A(y) + c_1^e c_1^q C(y), \nonumber\\
  & \tilde T_1^q(y) = c_1^e c_3^q A(y) + c_3^e c_1^q C(y),
\label{eq:T0T1}
\end{align}
to simplify the expressions.
$T_i^q(y)$'s and $\tilde T_i^q(y)$'s are related to space reflection even and odd structure respectively in the cross section.
For EM interaction, it requires $c_3^{e/q} = 0$ and $c_1^{e/q} = 1$. In this case, only $T_0^q(y)$ and $T_1^q(y)$ are left, and $T_0^q(y)=A(y)$, $T_1^q(y)=C(y)$. For the interference terms, we need to set $c_3^{e/q}=c_A^{e/q}$ and $c_1^{e/q}=c_V^{e/q}$.
The kinematic factors are also different.
To make it transparent, we can get the EM and interference cross sections by replacing the parameters in the weak interaction cross section according to Tab.~\ref{tab:replacing}

\begin{table}
\renewcommand\arraystretch{1.5}
\begin{tabular}{c|c|c|c}
\hline
~~~~  & $A_r$   & $L^{\mu\nu}_r$   & $W^{\mu\nu}_r$  \\ \hline
~ $ZZ$~ & $\chi$ & $c_1^e,~c_3^e$ & $c_1^q,~c_3^q$  \\
$\gamma Z$  & ~$\chi\to \chi_{int}$ ~& ~~$c_1^e\to c_V^e,~c_3^e\to c_A^e$~~ &~~ $c_1^q\to c_V^q,~c_3^q\to c_A^q$~~ \\
$\gamma\gamma$ & $\chi\to e_q^2$  & $c_1^e\to 1,~c_3^e\to 0$  & $c_1^q\to 1,~c_3^q\to 0$ \\ \hline
\end{tabular}
\caption{Relations of kinematic factors between weak, EM and interference interactions.}
\label{tab:replacing}
\end{table}

%\subsection{Twist-3 cross section}
Similarly, substituting the twist-3 hadronic tensor in Eq. (\ref{f:Wt3munu}) and the leptonic tensor into Eq.~(\ref{f:intcross}) yields the twist-3 cross section. It is given by
\begin{widetext}
\begin{align}
  \frac{d\sigma_{t3}^{ZZ}}{dx dy d\psi d^2 k_\perp^\prime} =& -\frac{\alpha_{\rm{em}}^2 \chi }{y Q^2}2x\kappa_M \Biggl\{ k_{\perp M}\cos\varphi \bigl(T_2^q(y)-\lambda_e \tilde T_2^q(y)\bigr)f^\perp+k_{\perp M}\sin\varphi \bigl(\tilde T_3^q(y)-\lambda_e T_3^q(y)\bigr)g^\perp \nonumber\\
%%%%%%%%%%%%%%%%%%%%%%%%%
  &+\lambda_h k_{\perp M}\Big[\sin\varphi \bigl( T_2^q(y)-\lambda_e \tilde T_2^q(y)\bigr)f^\perp_L - \cos\varphi \bigl( \tilde T_3^q(y)-\lambda_e T_3^q(y)\bigr)g_L^\perp\Big] \nonumber\\
  &+|S_T|\Big[\sin\varphi_S \bigl(  T_2^q(y)-\lambda_e \tilde T_2^q(y)\bigr)f_T -\cos\varphi_S \bigl( \tilde T_3^q(y)-\lambda_e T_3^q(y)\bigr)g_T  \nonumber\\
  &\quad +\sin(2\varphi-\varphi_S) \bigl( T_2^q(y)-\lambda_e \tilde T_2^q(y)\bigr)\frac{k_{\perp M}^2}{2}f^\perp_T -\cos(2\varphi-\varphi_S) \bigl( \tilde T_3^q(y)-\lambda_e T_3^q(y)\bigr)\frac{k_{\perp M}^2}{2}g^\perp_T \Big]
 \Biggr\},
\end{align}
\end{widetext}
where $\kappa_M=M/Q$ is a twist suppression factor.
We have also defined
\begin{align}
  & T_2^q(y) = c_1^e c_1^q B(y) + c_3^e c_3^q D(y), \nonumber\\
  & \tilde T_2^q(y) = c_3^e c_1^q B(y) + c_1^e c_3^q D(y), \nonumber\\
  & T_3^q(y) = c_3^e c_3^q B(y) + c_1^e c_1^q D(y), \nonumber\\
  & \tilde T_3^q(y) = c_1^e c_3^q B(y) + c_3^e c_1^q D(y).
\label{eq:T2T3}
\end{align}

It is also straightforward to obtain the interference and EM differential cross sections by doing the corresponding replacements.
To further unify the notations, we define $T_{i,r}^q(y)$'s and $\tilde T_{i,r}^q(y)$'s with $r=ZZ$, $\gamma Z$ and $\gamma\gamma$. For the weak interaction, we have $T_{i,ZZ}^q(y)$'s and $\tilde T_{i,ZZ}^q(y)$'s defined as $T_{i}^q(y)$'s and $\tilde T_{i}^q(y)$'s given in Eqs.~(\ref{eq:T0T1}) and (\ref{eq:T2T3}) respectively.
For interference and EM parts, according to Tab.~\ref{tab:replacing}, we have:
\begin{align}
  & T_{0,\gamma Z}^q(y) = c_V^e c_V^q A(y) + c_A^e c_A^q C(y), \nonumber\\
  & \tilde T_{0,\gamma Z}^q(y) = c_A^e c_V^q A(y) + c_V^e c_A^q C(y), \nonumber\\
  & T_{1,\gamma Z}^q(y) = c_A^e c_A^q A(y) + c_V^e c_V^q C(y), \nonumber\\
  & \tilde T_{1,\gamma Z}^q(y) = c_V^e c_A^q A(y) + c_A^e c_V^q C(y), \nonumber\\
  & T_{2,\gamma Z}^q(y) = c_V^e c_V^q B(y) + c_A^e c_A^q D(y), \nonumber\\
  & \tilde T_{2,\gamma Z}^q(y) = c_A^e c_V^q B(y) + c_V^e c_A^q D(y), \nonumber\\
  & T_{3,\gamma Z}^q(y) = c_A^e c_A^q B(y) + c_V^e c_V^q D(y), \nonumber\\
  & \tilde T_{3,\gamma Z}^q(y) = c_V^e c_A^q B(y) + c_A^e c_V^q D(y),
\label{eq:Ts-gammaZ}
\end{align}
and
\begin{align}
  & T_{0,\gamma\gamma}^q(y) = A(y), \quad \tilde T_{0,\gamma\gamma}^q(y) = 0 \nonumber\\
  & T_{1,\gamma\gamma}^q(y) = C(y), \quad \tilde T_{1,\gamma\gamma}^q(y) = 0 \nonumber\\
  & T_{2,\gamma\gamma}^q(y) = B(y), \quad \tilde T_{2,\gamma\gamma}^q(y) = 0 \nonumber\\
  & T_{3,\gamma\gamma}^q(y) = D(y), \quad \tilde T_{3,\gamma\gamma}^q(y) = 0.
\label{eq:Ts-gammagamma}
\end{align}
We see that only half of the terms will survive if only EM interaction is considered. This is because parity is conserved in EM interactions.

\subsection{The intrinsic asymmetry}

Most of the discussions based on the differential cross section are about the azimuthal and spin asymmetries.  They are important for understanding the TMDs and/or nucleon structures. In this part, we introduce a new quantity, named intrinsic asymmetry, to explore the transverse momentum distribution of  the quark in a nucleon.

\begin{figure}
\centering
\includegraphics[width= 0.4\linewidth]{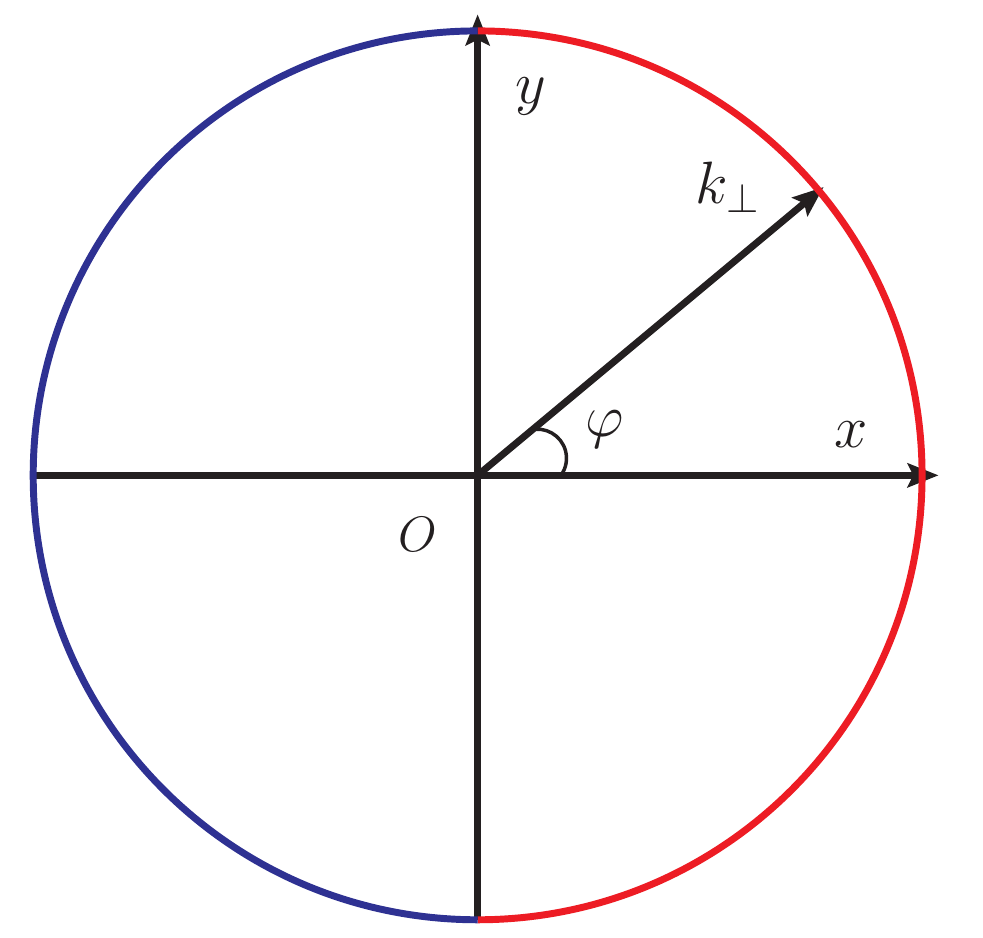}
\caption{The quark (jet) transverse momentum in the x-y plane. The difference of the red hemisphere and the blue one gives the imbalance of the transverse momentum distribution.}
\label{fig:asy}
\end{figure}

In the $\gamma^* N$ collinear frame introduced before, the transverse momentum of the incident quark (jet) is in the x-y plane. It can be decomposed as
\begin{align}
 & k_\perp^{x}=k_\perp \cos\varphi, \\
 & k_\perp^{y}=k_\perp \sin\varphi.
\end{align}
Therefore, it is possible to explore the imbalance of the momentum in the $x$-direction, i.e., $k_\perp^x (+x)-k_\perp^x (-x)$, see Fig. \ref{fig:asy}. This imbalance would be induced by the intrinsic transverse momentum of the quark in the
nucleon (we do not consider the contributions from gluons). To explore this imbalance, we can define the asymmetry as
\begin{align}
 %A^x = \frac{\int_{-\pi/2}^{\pi/2} d\varphi ~d\tilde{\sigma} -\int_{\pi/2}^{3\pi/2} d\varphi~ d\tilde{\sigma}}{\int_{-\pi/2}^{3\pi/2} d\tilde{\sigma}_{U,U} d\varphi} \label{f:akx}
 A^x = \frac{\int^{} d\varphi ~d\tilde{\sigma}(+x) -\int^{} d\varphi~ d\tilde{\sigma}(-x)}{\int d\tilde{\sigma}_{U,U} d\varphi} \label{f:akx}
\end{align}
for $S_T$-independent case.
%The superscript $(\pm)x$ denotes the (positive and negative) $x$-direction.
Here $d\tilde \sigma$ is used to denote $\frac{d\sigma}{dx dy d\psi d^2 k_\perp^\prime}$. Subscript $U,U$ denotes the unpolarized cross section. The  sum of the differential cross section for EM, weak and interference terms is understood. For the $S_T$-dependent case, we define
\begin{align}
% A^x=\frac{\int_{-\pi/2}^{\pi/2} d\varphi \int_{-\pi/2}^{\pi/2} d\varphi_S d\tilde{\sigma}-\int_{\pi/2}^{3\pi/2} d\varphi \int_{-\pi/2}^{\pi/2} d\varphi_S d\tilde{\sigma}}{\int_{-\pi/2}^{3\pi/2} d\varphi \int_{-\pi/2}^{\pi/2}d\varphi_S d\tilde{\sigma}_{U,U}}. \label{f:aksx}
 A^x=\frac{\int^{} d\varphi \int d\varphi_S d\tilde{\sigma}(+x)-\int^{} d\varphi \int d\varphi_S d\tilde{\sigma}(-x)}{\int d\varphi \int d\varphi_S d\tilde{\sigma}_{U,U}}. \label{f:aksx}
\end{align}
One notes that, in Eqs. (\ref{f:akx}) and (\ref{f:aksx}), we only introduced the asymmetries in the $x$-direction. Asymmetries in the $y$-direction can be defined in the similar way. We do not show them here for simplicity.

According to our definition, we find that these asymmetries do not vanish, at least formally. For the $S_T$-independent asymmetries, we have
\begin{align}
 & A_{U,U}^x = - \frac{4x\kappa_M k_{\perp M}}{\pi} \frac{\chi T^q_2(y)f^\perp}{\chi T^q_0(y) f_1}, \label{f:auux} \\
 & A_{U,U}^y = - \frac{4x\kappa_M k_{\perp M}}{\pi} \frac{\chi \tilde T^q_3(y)g^\perp}{\chi T^q_0(y) f_1}, \label{f:auuy} \\
 & A_{U,L}^x =  \frac{4x\kappa_M k_{\perp M}}{\pi} \frac{\chi \tilde T^q_3(y)g_L^\perp}{\chi T^q_0(y) f_1}, \label{f:aulx} \\
 & A_{U,L}^y = - \frac{4x\kappa_M k_{\perp M}}{\pi} \frac{\chi T^q_2(y)f_L^\perp}{\chi T^q_0(y) f_1}, \label{f:auly} \\
%%%%%%%%%%%%%%%%%%%%%%%
 & A_{L,U}^x = \frac{4x\kappa_M k_{\perp M}}{\pi} \frac{\chi \tilde T^q_2(y)f^\perp}{\chi T^q_0(y) f_1}, \label{f:alux} \\
 & A_{L,U}^y = \frac{4x\kappa_M k_{\perp M}}{\pi} \frac{\chi T^q_3(y)g^\perp}{\chi T^q_0(y) f_1}, \label{f:aluy} \\
 & A_{L,L}^x = -\frac{4x\kappa_M k_{\perp M}}{\pi} \frac{\chi T^q_3(y)g_L^\perp}{\chi T^q_0(y) f_1}, \label{f:allx} \\
 & A_{L,L}^y = \frac{4x\kappa_M k_{\perp M}}{\pi} \frac{\chi \tilde T^q_2(y)f_L^\perp}{\chi T^q_0(y) f_1}. \label{f:ally}
\end{align}
We can see that they are twist-3 effects and are suppressed by a factor $\kappa_M$. There are four $S_T$-dependent asymmetries which correspond to the leading twist effects.
\begin{align}
 & A_{U,T}^x =- \frac{4 k_{\perp M}}{\pi^2} \frac{\chi \tilde T^q_1(y)g_{1T}^\perp}{\chi T^q_0(y) f_1}, \label{f:autx} \\
 & A_{U,T}^y = \frac{4 k_{\perp M}}{\pi^2} \frac{\chi T^q_0(y)f_{1T}^\perp}{\chi T^q_0(y) f_1}, \label{f:auty} \\
 & A_{L,T}^x = \frac{4 k_{\perp M}}{\pi^2} \frac{\chi T^q_1(y)g_{1T}^\perp}{\chi T^q_0(y) f_1}, \label{f:altx} \\
 & A_{L,T}^y = - \frac{4 k_{\perp M}}{\pi^2} \frac{\chi \tilde T^q_0(y)f_{1T}^\perp}{\chi T^q_0(y) f_1}. \label{f:alty}
\end{align}
We see that $A_{U,T}^y$ and $A_{L,T}^y$ are determined by the Sivers function $f_{1T}^\perp$ \cite{Sivers:1989cc,Sivers:1990fh}.
We note again that only weak interaction results are shown in Eqs. (\ref{f:auux})-(\ref{f:alty}). For the complete results, EM and interference interactions should be included.

If only the EM interaction is considered, we are left with the following asymmetries,
\begin{align}
 & A_{U,U}^{\gamma\gamma,x} = - \frac{4x\kappa_M k_{\perp M}}{\pi} \frac{e_q^2 B(y)f^\perp}{e_q^2 A(y) f_1}, \label{f:auuxgg} \\
 & A_{U,L}^{\gamma\gamma,y} = - \frac{4x\kappa_M k_{\perp M}}{\pi} \frac{e_q^2 B(y)f_L^\perp}{e_q^2 A(y) f_1}, \label{f:aulygg} \\
%%%%%%%%%%%%%%%%%%%%%%%
 & A_{L,U}^{\gamma\gamma,y} = \frac{4x\kappa_M k_{\perp M}}{\pi} \frac{e_q^2 D(y)g^\perp}{e_q^2 A(y) f_1}, \label{f:aluygg} \\
 & A_{L,L}^{\gamma\gamma,x} = -\frac{4x\kappa_M k_{\perp M}}{\pi} \frac{e_q^2 D(y)g_L^\perp}{e_q^2 A(y) f_1}, \label{f:allxgg}\\
%%%%%%%%%%%%%%%%%%%%
 & A_{U,T}^{\gamma\gamma,y} = \frac{4 k_{\perp M}}{\pi^2} \frac{e_q^2 A(y)f_{1T}^\perp}{e_q^2 A(y) f_1}, \label{f:autygg} \\
 & A_{L,T}^{\gamma\gamma,x} = \frac{4 k_{\perp M}}{\pi^2} \frac{e_q^2 C(y)g_{1T}^\perp}{e_q^2 A(y) f_1}. \label{f:altxgg}
\end{align}
However, Eqs. (\ref{f:auuxgg})-(\ref{f:altxgg}) cannot give any information about the electroweak couplings.
To determine these couplings, we still need to study the asymmetries from both the weak and EM interactions. 

To have an intuitive impression of the intrinsic asymmetries shown above, we present the numerical values of $A_{U,U}^x$ and $A_{L,U}^x$ in Figs. \ref{fig:Axuuy-u} to \ref{fig:Axuuk-d}, respectively. We take the Gaussian ansatz for the TMDs, i.e.,
\begin{align}
 & f_1(x, k_\perp) =\frac{1}{\pi \Delta^2} f_1(x) e^{-\vec k_\perp^2/\Delta^2}, \\
 & f^\perp(x, k_\perp) =\frac{1}{\pi \Delta^2x}f_1(x) e^{-\vec k_\perp^2/\Delta^2},
\end{align}
where $f_1(x)$ are taken from CT14 \cite{Schmidt:2015zda}. In order to determine $f^\perp(x, k_\perp)$, we have used the Wandzura-Wilczek approximation (neglecting quark-gluon-quark correlation function, $g=0$) \cite{Mulders:1995dh,Bacchetta:2006tn}. Only the $u$ and $d$ quarks are taken into account. Figures \ref{fig:Axuuy-u} and \ref{fig:Axuuk-u} show the results at $\Delta_u=0.5$~GeV while Figs. \ref{fig:Axuuy-d} and \ref{fig:Axuuk-d} show the results at $\Delta_d=0.5$~GeV.

A few remarks are listed as follows:
\begin{itemize}
 \item $A_{L,U}^x$ and  $A_{U,U}^x$ have the similar behaviors. However, asymmetry $A_{L,U}^x$ is two or three orders of magnitude smaller than $A_{U,U}^x$. From Eq. (\ref{f:alux}) we know $A_{L,U}^x$ is a parity violating effect or an effect of the weak interaction. It should be the same order of magnitude as parity violation in standard model.

 \item A surprising consequence is that the magnitude of $A_{L,U}^x$ varies according to the difference between $\Delta_u$ and $\Delta_d$. If the difference is zero, i.e., $\Delta_d=\Delta_u$, $A_{L,U}^x$ has the same behaviors as $A_{U,U}^x$.

 \item Furthermore, the asymmetry $A_{U,U}^x$ is insensitive to the factor $\Delta$ and decrease with respect to the energy, while $A_{L,U}^x$ is sensitive to $\Delta$ and increase with the energy.
\end{itemize}

%We note that Wandzura-Wilczek approximation is not physical due to the ignorance of the quark-gluon-quark corelation function. This will break the gauge invariance.

\begin{widetext}

\begin{figure}
\centering
\includegraphics[width= 0.4\linewidth]{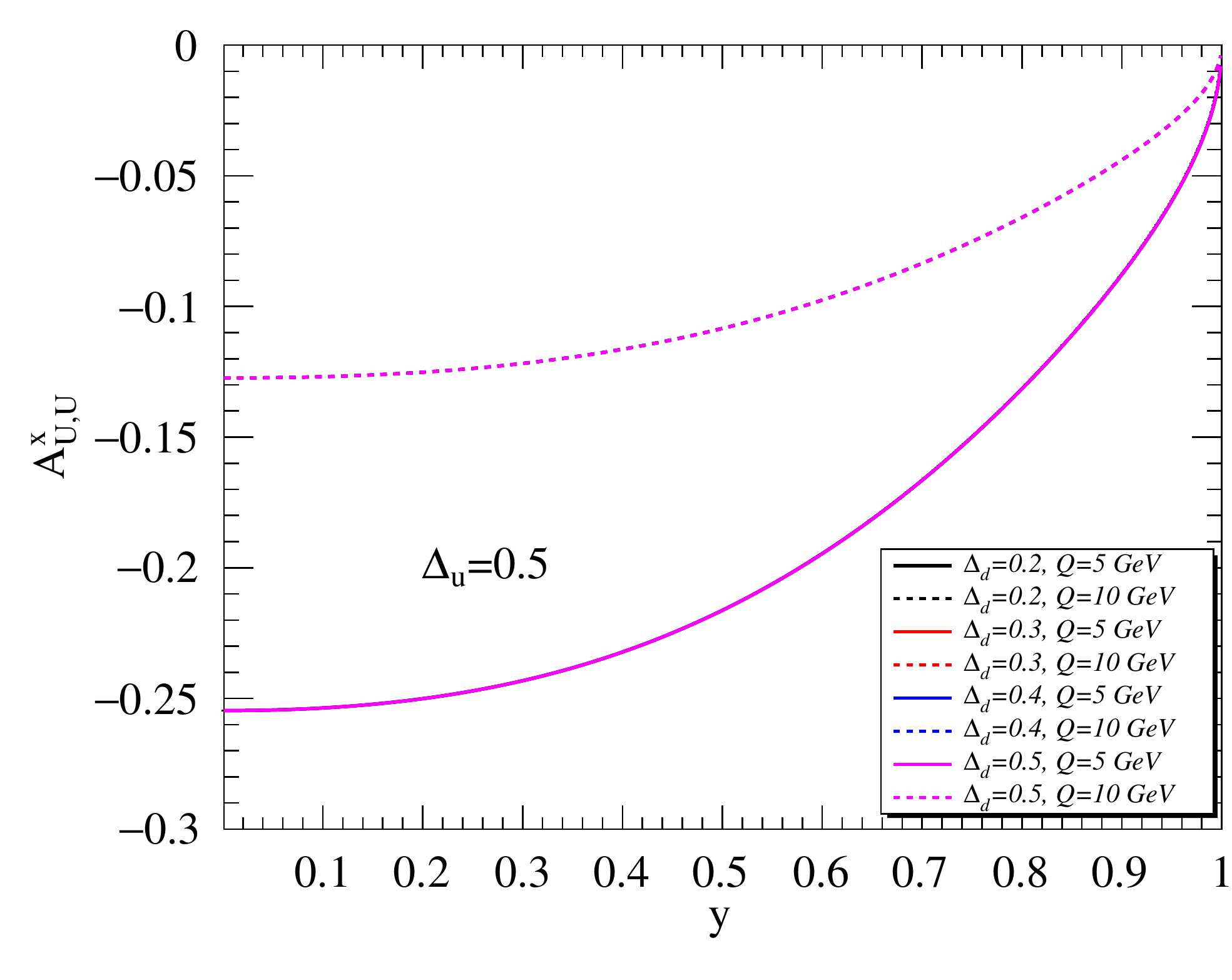}
\quad \quad \quad
\includegraphics[width= 0.4\linewidth]{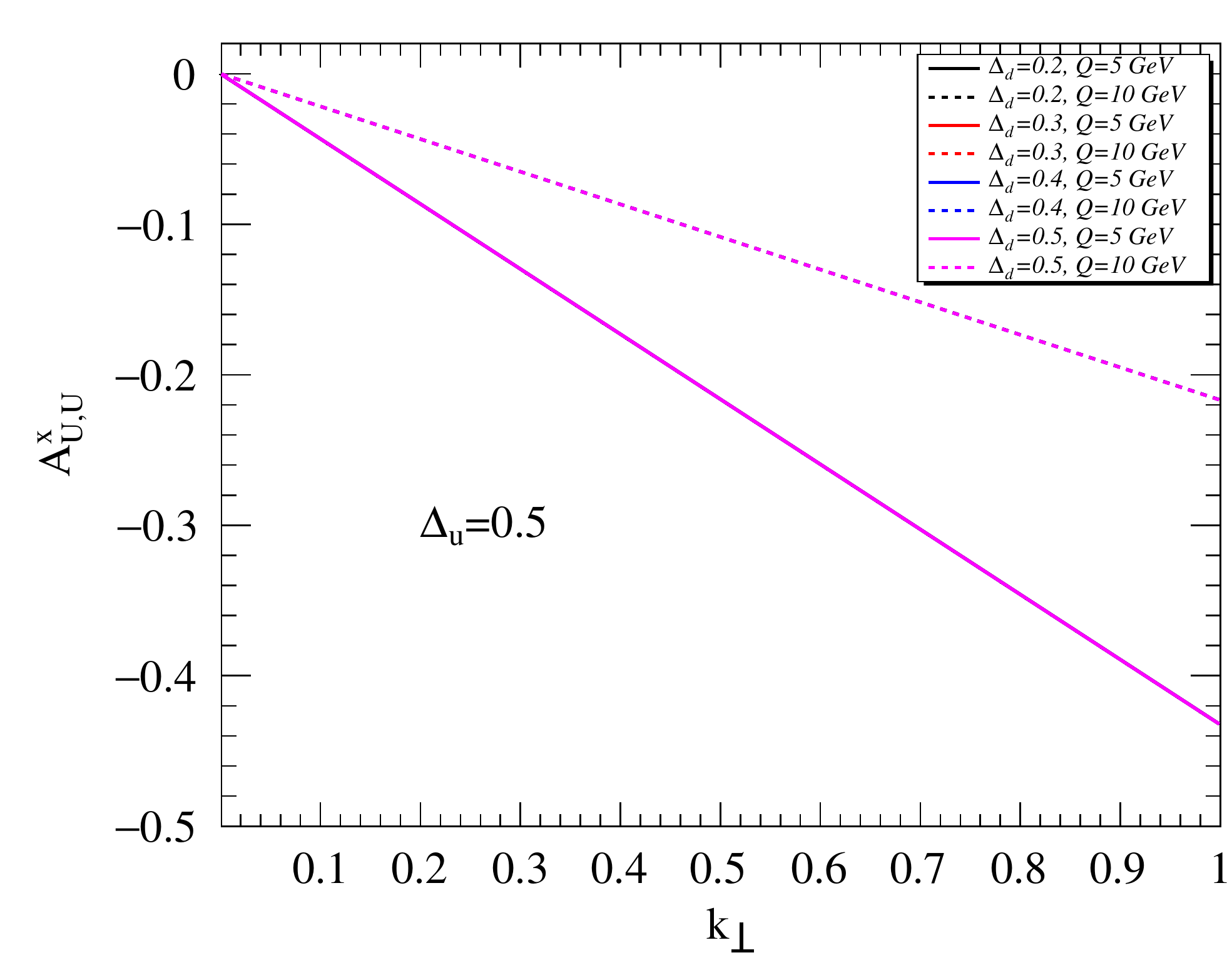}
\caption{The intrinsic asymmetry $A_{U,U}^x$ with respect to $y$ (left) and $k_\perp$ (right). The solid lines show the asymmetry at 5~GeV while the dashed lines show the asymmetry at $Q=$10~GeV. Here $\Delta_u=0.5$~GeV while $\Delta_d$ runs from $0.2$ to $0.5$~GeV. }
\label{fig:Axuuy-u}
\end{figure}
\begin{figure}
\centering
\includegraphics[width= 0.4\linewidth]{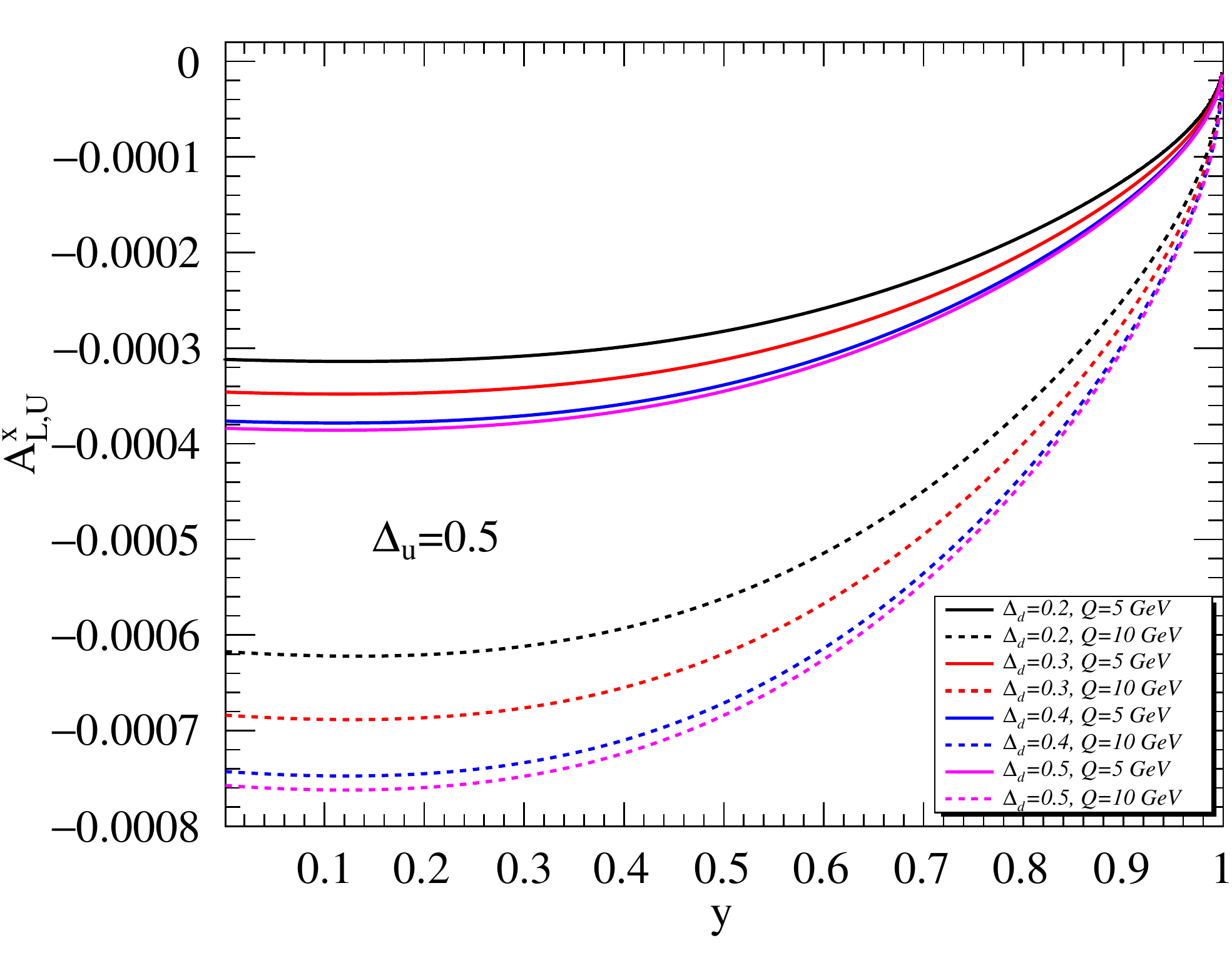}
\quad \quad \quad
\includegraphics[width= 0.4\linewidth]{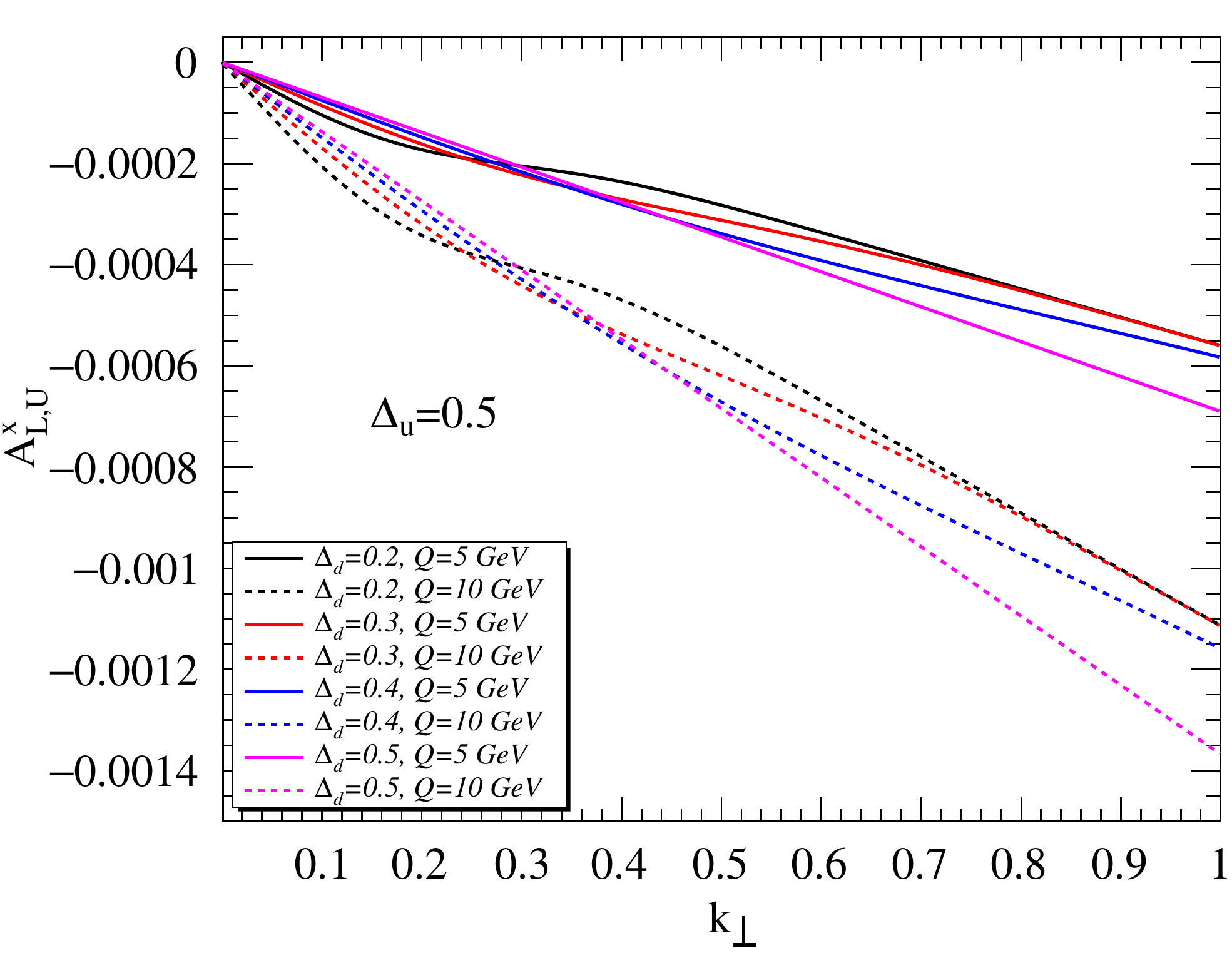}
\caption{The intrinsic asymmetry $A_{L,U}^x$ with respect to $y$ (left) and $k_\perp$ (right). The solid lines show the asymmetry at 5~GeV while the dashed lines show the asymmetry at $Q=$10~GeV. Here $\Delta_u=0.5$~GeV while $\Delta_d$ runs from $0.2$ to $0.5$~GeV.}
\label{fig:Axuuk-u}
\end{figure}
\begin{figure}
\centering
\includegraphics[width= 0.4\linewidth]{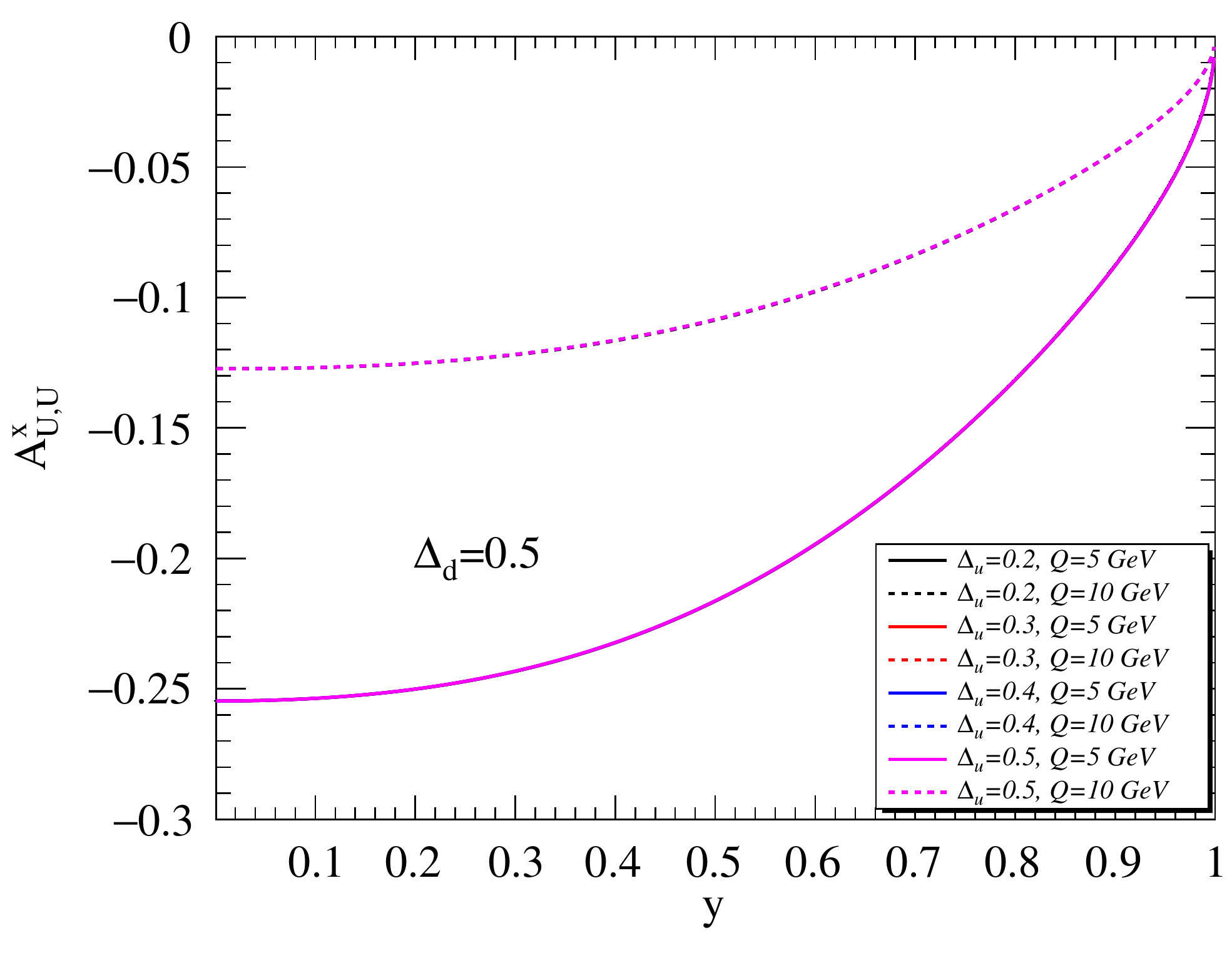}
\quad \quad \quad
\includegraphics[width= 0.4\linewidth]{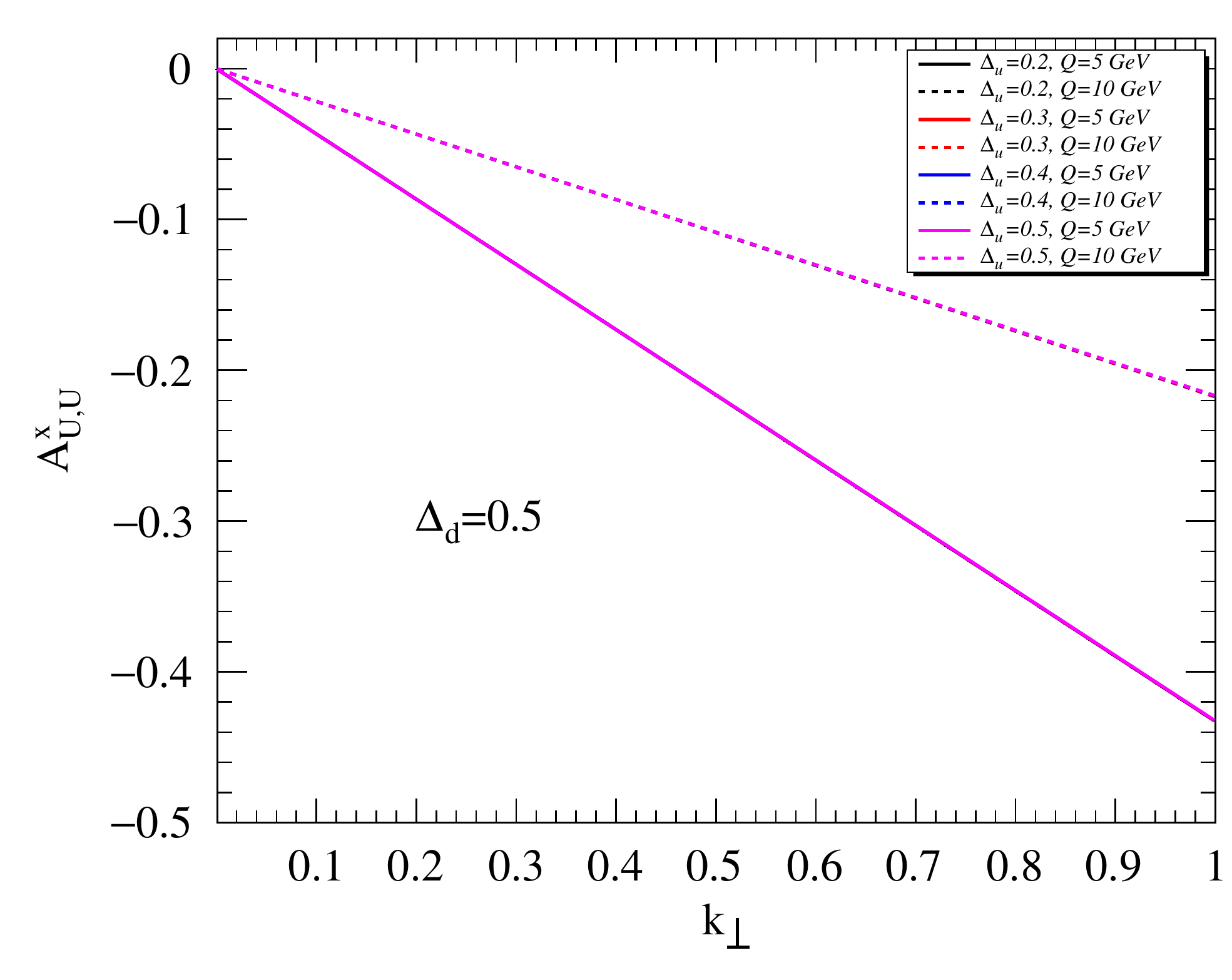}
\caption{The intrinsic asymmetry $A_{U,U}^x$ with respect to $y$ (left) and $k_\perp$ (right). The solid lines show the asymmetry at 5~GeV while the dashed lines show the asymmetry at $Q=$10~GeV. Here $\Delta_u=0.5$~GeV while $\Delta_d$ runs from $0.2$ to $0.5$~GeV.}
\label{fig:Axuuy-d}
\end{figure}
\begin{figure}
\centering
\includegraphics[width= 0.4\linewidth]{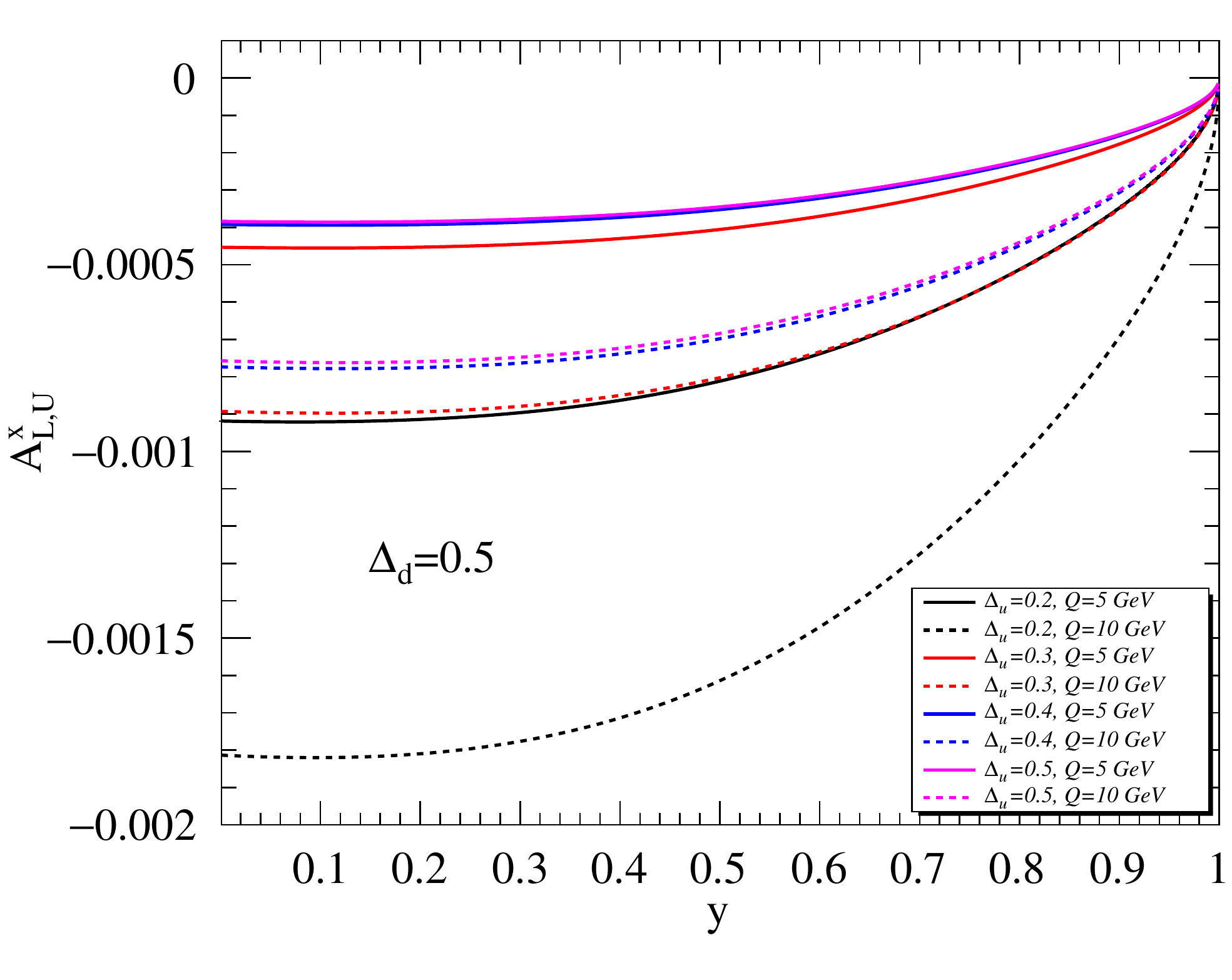}
\quad \quad \quad
\includegraphics[width= 0.4\linewidth]{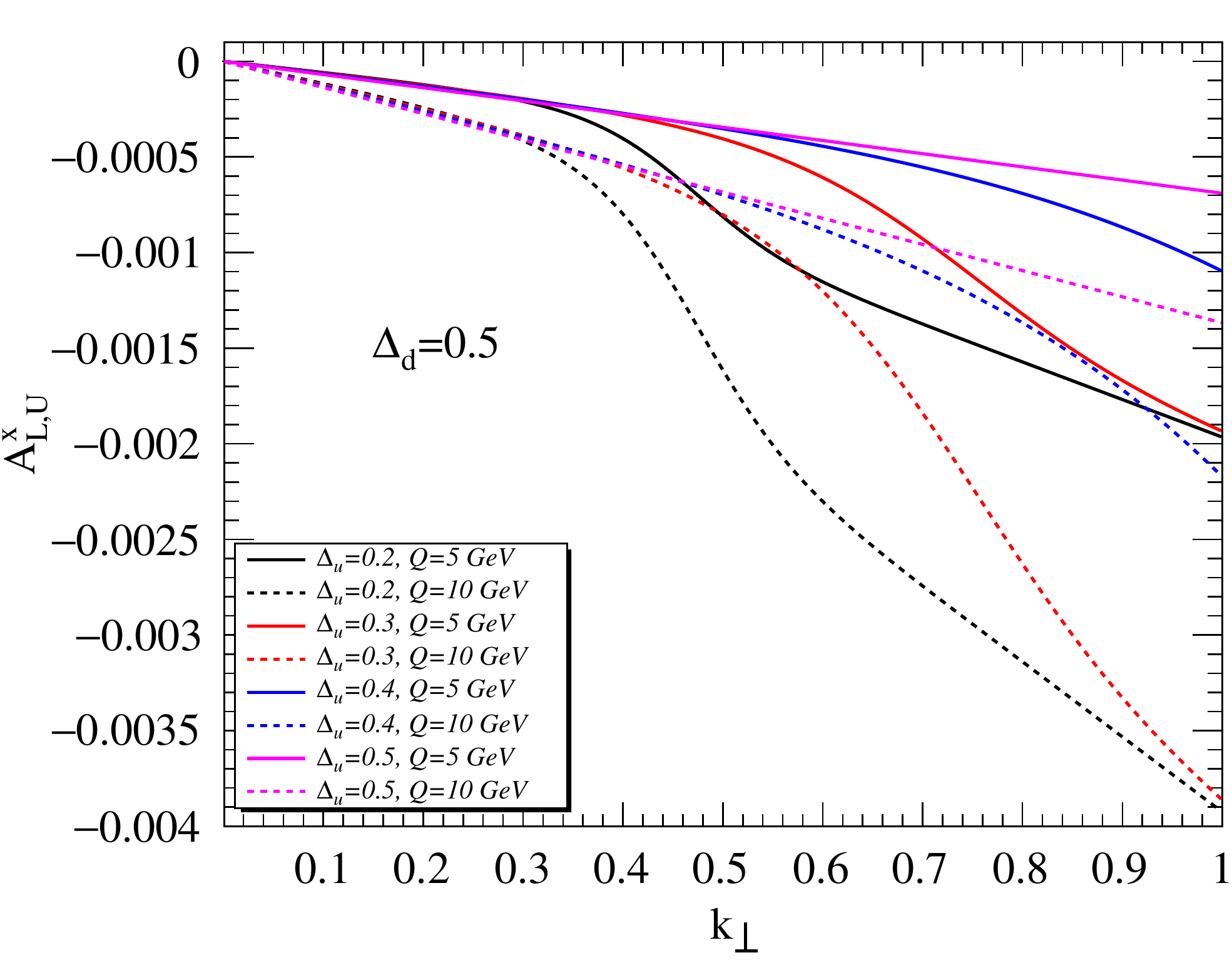}
\caption{The intrinsic asymmetry $A_{L,U}^x$ with respect to $y$ (left) and $k_\perp$ (right). The solid lines show the asymmetry at 5~GeV while the dashed lines show the asymmetry at $Q=$10~GeV. Here $\Delta_u=0.5$~GeV while $\Delta_d$ runs from $0.2$ to $0.5$~GeV. }
\label{fig:Axuuk-d}
\end{figure}

\end{widetext}

\section{Summary} \label{sec:summary}

In this paper, we consider the neutral current jet production SIDIS process and calculate the differential cross section of this process at tree level twist-3. The calculation includes the EM, weak and inference interactions. The initial electron is assumed to be polarized and then scattered off by a target particle with spin-1/2. After obtaining the differential cross section, we introduce the definition of the intrinsic asymmetry which is induced from the quark intrinsic transverse momentum. We obtain eight $S_T$-independent asymmetries and four $S_T$-dependent asymmetries with well definitions. We find that these asymmetries can be expressed in terms of the TMDs and the electroweak couplings. To have an intuitive impression of these intrinsic asymmetries, we present the numerical values of $A_{U,U}^x$ and $A_{L,U}^x$. A few observations are also shown in the last section. First of all, we find that $A_{L,U}^x$ is two or three orders of magnitude smaller than $A_{U,U}^x$, as it is a parity violating effect. Second, the magnitude of $A_{L,U}^x$ depends on the difference between $\Delta_u$ and $\Delta_d$. Third, $A_{U,U}^x$ is insensitive to the factor $\Delta$ and decrease with respect to the energy, while $A_{L,U}^x$ is sensitive to $\Delta$ and increase with the energy.
In a word, our calculations provide a set of new quantities for analyzing these corresponding TMDs and the electroweak couplings.  It is helpful to understand the hadronic weak interactions and strong interactions as well as the nucleon structures in the deeply inelastic scattering process simultaneously.

\section*{Acknowledgements}
The authors thank Zhe Zhang very much for his kind help. This work was supported by the Natural Science Foundation of Shandong Province (Grants No. ZR2021QA015 and ZR2021QA040).

\end{document}